\documentclass[graybox]{svmult}

\usepackage{mathptmx}       
\usepackage{helvet}         
\usepackage{courier}        
\usepackage{type1cm}        
\usepackage{makeidx}         
\usepackage{graphicx}        
\usepackage{multicol}        
\usepackage[bottom]{footmisc}

\usepackage{cite}
\usepackage{fancyvrb}        

\usepackage{amsmath}
\usepackage{amssymb}

\newcommand{\onehalf}{{\scriptstyle\frac{1}{2}}}
\newcommand{\N}{\mathbb{N}}

\makeindex             

\begin{document}

\title*{On a numerical approximation scheme for construction of the early exercise boundary for a class of nonlinear Black--Scholes equations}
\titlerunning{Numerical approximation scheme for construction of the early exercise boundary} 

\author{Daniel \v Sev\v covi\v c }

\institute{Daniel \v Sev\v covi\v c  \at 
Department of Applied Mathematics and Statistics,
Faculty of Mathematics, Physics \& Informatics, Comenius University, 
842 48 Bratislava, Slovak Republic, \email{sevcovic@fmph.uniba.sk}
}

\maketitle

\abstract*{
The purpose of this paper is to construct the early exercise boundary 
for a class of nonlinear Black--Scholes equations with a nonlinear volatility depending on the option price. We review a method how to transform the problem into a solution of a time depending nonlinear parabolic equation defined on a fixed domain. Results of numerical computation of the early exercise boundary for various nonlinear Black--Scholes equations are also presented. 
}

\abstract{
The purpose of this paper is to construct the early exercise boundary 
for a class of nonlinear Black--Scholes equations with a volatility function depending on the option price. We review and revisit a method how to transform the problem into a solution of a time depending nonlinear parabolic equation defined on a fixed domain. An example of numerical computation of the early exercise boundary for a nonlinear Black--Scholes equation is also presented. 
}

\section{Black--Scholes equations with a nonlinear volatility function}

The main purpose of this paper is to review and revisit the fixed domain transformation method adopted for solving a class of nonlinear Black--Scholes equations having the form: 
\begin{equation}
\frac{\partial V}{\partial t} +  (r-q) S\frac{\partial V}{\partial S} + \frac{1}{2}\sigma^2(S^2\partial^2_S V, S, T-t)  S^2 \frac{\partial^2 V}{\partial S^2} - r V =0, \ \ S>0,\  t\in(0,T). 
\label{doplnky-nonlinear-BS}
\end{equation}
A solution $V=V(S,t)$ can be identified with a price $V$ of the option contract in the future maturity time $T>0$ (e.g. call or put) where $S>0$ is the underlying asset value at the present time $t\in[0,T)$. Here, $r>0$ is the riskless interest rate, $q\ge 0$  is the dividend yield rate of the underlying asset. For American style of a call option, the free boundary problem consists in construction of the early exercise position $S_f=S_f(t)$ and the solution $V=V(S,t)$ to equation (\ref{doplnky-nonlinear-BS}) defined on the time dependent domain $0<S<S_f(t),\ 0<t<T$ (cf. Kwok \cite{doplnkyKw}). $V$ is subjected to the boundary conditions yielding $C^1$ smooth pasting of $V(S,t)$ and $V(S,T)$ at $S=S_f(t)$:
\begin{equation}
V(0,t) = 0\,,\ \ V(S_f(t), t)= S_f(t)-E\,, \ \ \partial_S  V (S_f(t),t) = 1\,,
\label{doplnky-nonlinear-bccall}
\end{equation}
and the terminal pay-off condition at expiry $t=T,$
\begin{equation}
V(S,T)=(S-E)^+,
\label{doplnky-nonlinear-tccall}
\end{equation}
where $E>0$ is the exercise price. 

We briefly mention a motivation for studying the nonlinear Black--Scholes equation having the form of (\ref{doplnky-nonlinear-BS}). Such equations with a volatility function $\sigma(S^2\partial^2_S V, S, T-t)$ arise from  modeling the option prices by taking into account nontrivial transaction costs (cf. Leland \cite{doplnkyLe}, Hoggard {\em et al.} \cite{HWW}, Avellaneda and  Paras  \cite{doplnkyAP}), market feedbacks and effects due to large traders choosing given stock-trading  strategies (Frey \cite{Frey2000}, Frey and Patie \cite{FP}, Frey and Stremme \cite{FS}, During {\em et al.}\cite{doplnkyDFJ}, Sch\"onbucher and  Wilmott \cite{SW}), the risk adjusted pricing methodology model due to Kratka \cite{doplnkyKr} and its modification developed by Janda\v{c}ka and \v{S}ev\v{c}ovi\v{c} \cite{doplnkyJS,doplnkySe2}). As an example for application of the numerical method, we consider on a nonlinear model taking into account imperfect replication and investor's preferences which has been proposed by Barles and Soner in \cite{doplnkyBaSo}. If investor's preferences are characterized by an exponential utility function they  derived a nonlinear Black--Scholes equation with the volatility function $\sigma$ given by
\begin{equation}
\sigma^2(S^2\partial^2_S V, S, \tau) 
= \hat\sigma^2 \left(1+\Psi(a^2 e^{r \tau} S^2\partial^2_S V)\right).
\label{doplnky-c-barles}
\end{equation}
Here $\hat\sigma^2>0$ is a constant historical volatility of the asset price returns, $\Psi$ is the unique solution to the ODE: $\Psi^\prime(x) = (\Psi(x)+1)/(2 \sqrt{x\Psi(x)} -x), \Psi(0)=0$ and $a\ge 0$ is a constant depending transaction costs and investor's risk aversion parameter (see \cite{doplnkyBaSo} for details). The function $\Psi$ satisfies:  $\Psi(x)=O(x^\frac{1}{3})$ for  $x\to 0$ and $\Psi(x)=O(x)$ for $x\to\infty$. For practical purposes, the solution $\Psi(x)$ can be constructed from an implicit equation obtained in \cite{Comp}.

We revisit an iterative numerical algorithm for solving the free boundary problem (\ref{doplnky-nonlinear-BS})--(\ref{doplnky-nonlinear-bccall}) developed by  \v{S}ev\v{c}ovi\v{c} in \cite{doplnkySe2}. The key idea of this method  consists in transformation of the free boundary problem into a semilinear  parabolic equation  defined  on a fixed spatial domain coupled with a nonlocal algebraic constraint equation for the free boundary position. This method has been analyzed and utilized in a series of papers  \cite{doplnkyAE1,doplnkyAE2,doplnkySSC,doplnkySe,doplnkySe2,doplnkySe3} by Ehrhardt and Ankudinova and the author. The disadvantage of the original method consists in the necessity of solving an algebraic constraint equation. In this approach, highly accurate evaluation of the derivative of a solution at one point entering the algebraic constraint is needed (cf. \cite{doplnkySe2}). In this note, we present a new efficient way how to overcome this difficulty by considering  an equivalent integrated form of the algebraic constraint. We also present results of numerical calculation of the free boundary position for the Barles and Soner nonlinear extension of the Black--Scholes model. 

\section{Fixed domain transformation of the free boundary problem}

We recall the method how to transform the free boundary problem (\ref{doplnky-nonlinear-BS})--(\ref{doplnky-nonlinear-tccall}) into a form of a nonlinear parabolic equation defined on a fixed domain and satisfying a nonlocal algebraic constraint equation developed by the author in \cite{doplnkySe2}. It is based on the following change of independent variables and the transformed function $\Pi=\Pi(x,\tau)$ defined as follows:
\begin{equation}
\tau= T-t, \quad x=\ln\left(\varrho(\tau)/S\right), \quad \Pi(x,\tau)= V(S,t) - S \partial_S V (S,t),
\label{doplnky-c-transformacia}
\end{equation}
where $\varrho(\tau)=S_f(T-\tau)$. Clearly,  $\tau\in(0,T)$ and $x\in(0,\infty)$ iff $S\in(0,S_f(t))$. The boundary value $x=0$ corresponds to the free boundary position $S=S_f(t)$ whereas $x =  +\infty$ corresponds to the default value $S=0$ of the underlying asset. Under the structural assumption 
\[
0<q \le r
\] 
made on the interest and dividend yield rates and following derivation of the equation for $\Pi$, it turns out that the function $\Pi$ and the free boundary position $\varrho$ satisfy the following system of parabolic equation (\ref{doplnky-c-tBS}) with algebraic constraint (\ref{doplnky-c-ro}): 
\begin{eqnarray}
&&\frac{\partial\Pi}{\partial \tau}  + \left( b(\tau) -\frac{\sigma^2}{2} \right)\frac{\partial \Pi}{\partial x}
- \frac12  \frac{\partial}{\partial x}\left( \sigma^2 \frac{\partial \Pi}{\partial x}\right) + r \Pi=0\,, \label{doplnky-c-tBS}
\\
&&\Pi(0, \tau) = -E\,, \ \  \Pi(+\infty, \tau) =0\,,\  x>0\,, \tau\in(0,T)\,,
\nonumber\\
&&\Pi(x,0)= \left\{
\begin{array}{lll}
-E & \quad\hbox{for} \ x<\ln(r/q),\hfil
\\
\ \ \  0, & \quad\hbox{otherwise\,,}\hfil
\end{array}
\right. \nonumber 
\\
&& \varrho(\tau)={r E\over q} + {\sigma^2(\partial_x\Pi(0,\tau), \varrho(\tau), \tau )\over 2q}
\frac{\partial \Pi}{\partial x}(0,\tau)\,,\quad \hbox{with }\ \varrho(0)={rE\over q},
\label{doplnky-c-ro}
\end{eqnarray}
where
$\sigma^2=\sigma^2(\partial_x\Pi(x,\tau), \varrho(\tau)e^{-x},\tau )\,,\ 
b(\tau)= {\dot\varrho(\tau)\over\varrho(\tau)} + r - q$ (cf. \cite{doplnkySe2}). Notice that equation (\ref{doplnky-c-ro}) is not quite appropriate for construction of a robust numerical approximation scheme since any small inaccuracy in approximation of the value $\partial_x\Pi(0,\tau)$ is immediately transferred in to the entire computational domain $x\in(0,\infty)$ through the free boundary function $\varrho(\tau)$ entering (\ref{doplnky-c-tBS}). Instead of (\ref{doplnky-c-ro}), we present a new equivalent integrated equation for the free boundary position $\varrho(\tau)$. Indeed, integrating the governing equation (\ref{doplnky-c-tBS}) for $x\in(0,\infty)$ taking into account the boundary conditions $\Pi(0,\tau)=-E, \Pi(\infty,\tau)=0$ (and consequently $\partial_x\Pi(\infty,\tau)=0$), we obtain the following spatially integrated form of the algebraic constraint:
\begin{eqnarray}
\label{doplnky-nonlin-alternative}
&& \frac{d}{d\tau}\Big( E \ln\varrho(\tau) + \int_0^\infty \hskip-2truemm\Pi(x,\tau) dx \Big) 
 + q\varrho(\tau) - q E \nonumber \\
&& + \int_0^\infty \left(
- \frac{1}{2} \sigma^2( \partial_x\Pi(x,\tau), \varrho(\tau) e^{-x}, \tau) \frac{\partial \Pi}{\partial x}(x,\tau) + r  \Pi(x,\tau) 
\right) dx =0.
\end{eqnarray}

\section{Numerical scheme based on operator splitting technique}

The idea of the iterative numerical algorithm is based on the original numerical discretization scheme proposed by the author in \cite{doplnkySe2}. We modify this method by considering the alternative integrated form (\ref{doplnky-nonlin-alternative}) of the constraint between $\Pi$ and $\varrho$. The spatial domain $x\in (0,\infty)$ is restricted to a finite interval of values $x\in (0,L)$ where $L>0$ is sufficiently large. For practical purposes one can take $L\approx 3$ (see \cite{doplnkySe2}). Let us denote by $k>0$ the time step, $k=T/m$ and by $h>0$ the spatial step, $h=L/n$ where $m, n\in \N$ stand for the number of time and space discretization steps, respectively.  We denote by $\Pi_i^j$ an approximation of $\Pi( x_i, \tau_j),$ $\varrho^j \approx \varrho(\tau_j),$ $b^j \approx b(\tau_j)$ where $x_i=i h, \tau_j=j k$. We furthermore denote by $\Pi^j$ the vector  $\Pi^j= \{ \Pi_i^j, i=1,\dots, n\}$. We approximate the value of the  volatility $\sigma$ at  the node $(x_i,\tau_j)$ by the finite difference approximation as follows:
\[
\sigma_i^j =\sigma((\Pi_{i+1}^j - \Pi_i^j)/h, \varrho^j e^{-x_i}, \tau_j)\,.
\]
We set $\Pi_i^0(x)=\Pi(x_i,0)$. Next, following the idea of the operator splitting method discussed in \cite{doplnkySe2}, we decompose the above problem into two parts - a convection part and a diffusive part by introducing 
an auxiliary intermediate step $\Pi^{j-\onehalf}$. Our discretization of equations (\ref{doplnky-nonlin-alternative}) and (\ref{doplnky-c-tBS}) reads as follows:

({\it Integrated form of the algebraic part})
\begin{equation}
\label{doplnky-c-eq-ro-integral}
E \ln \varrho^j = E \ln \varrho^{j-1} + I_0(\Pi^{j-1})
- I_0(\Pi^{j}) + k \left( q E - q\varrho^{j} -  I_1(\varrho^{j}, \Pi^j) \right),
\end{equation}
where $I_0(\Pi)$ stands for numerical trapezoid quadrature of the integral $\int_0^\infty  \Pi(\xi) d\xi$ whereas
$I_1(\varrho^{j},\Pi)$ is a trapezoid quadrature of the second integral in (\ref{doplnky-nonlin-alternative}), i.e. 
\[
I_1(\varrho^{j},\Pi)\approx
\int_0^\infty \left(- \frac12 \sigma^2( \partial_x\Pi(x), \varrho^j e^{-x}, \tau_j) 
\frac{\partial\Pi}{\partial x}(x)  + r \Pi(x) \right)dx\,.
\]

({\it Convective part})
\begin{equation}
\frac{\Pi^{j-\onehalf}-\Pi^{j-1}}{k} + b^j \frac{\partial}{\partial x} \Pi^{j-\onehalf} = 0\,,
\label{doplnky-c-convective}
\end{equation}

({\it Diffusive part})
\begin{equation}
\frac{\Pi^{j}-\Pi^{j-\onehalf}}{k}   - \frac{(\sigma^j)^2}{2}  \frac{\partial}{\partial x}\Pi^j
-\frac12 \frac{\partial}{\partial x} \left( (\sigma^j)^2 \frac{\partial}{\partial x}\Pi^j\right) +r \Pi^j = 0\,.
\label{doplnky-c-diffusion}
\end{equation}

The convective part can be approximated by an explicit solution to the transport equation $\partial_\tau\tilde\Pi + b(\tau) \partial_x\tilde\Pi =0$. Thus the spatial approximation $\Pi^{j-\onehalf}_i$ can be constructed from the formula
\begin{equation}
\Pi^{j-\onehalf}_i=\left\{ 
\begin{matrix}
\Pi^{j-1}(\xi_i) \hfill & \quad \hbox{if } \xi_i=x_i- \ln\varrho^j + \ln\varrho^{j-1} - (r-q)k>0\,, \hfill\cr 
-E, \hfill & \quad \hbox{otherwise,} \hfill
\end{matrix}
\right.
\label{doplnky-c-convective-discrete}
\end{equation}
where a piecewise linear interpolation  between discrete values $\Pi^{j-1}_i, i=0,1, ..., n,$ 
is being used to compute the value
$\Pi^{j-1}(x_i - \ln\varrho^j + \ln\varrho^{j-1} - (r-q)k)$.

The diffusive part can be solved numerically by means of finite differences. Using central finite difference approximation of the derivative $\partial_x\Pi^j$ we obtain
\begin{eqnarray}
\label{findif}
\frac{\Pi_i^j - \Pi_i^{j-\onehalf}}{k} + r\Pi_i^j
&-& \frac{(\sigma^j_i)^2}{2} \frac{\Pi_{i+1}^j-\Pi_{i-1}^j}{2h}
\nonumber \\
&-& \frac{1}{2 h} \left( 
(\sigma_i^j)^2 \frac{\Pi_{i+1}^j - \Pi_{i}^j}{h}
-
(\sigma_{i-1}^j)^2 \frac{\Pi_{i}^j - \Pi_{i-1}^j}{h}
\right)  =0\,.
\end{eqnarray}

Now, equations (\ref{doplnky-c-eq-ro-integral}), (\ref{doplnky-c-convective-discrete}) and  (\ref{findif}) can be rewritten in the operator form:
\[\varrho^j = {\mathcal F}(\Pi^j, \varrho^j),\quad
\Pi^{j-\onehalf} ={\mathcal T}(\Pi^j, \varrho^j),\quad
{\mathcal A}(\Pi^j, \varrho^j) \Pi^j = \Pi^{j-\onehalf},
\]
where 
${\mathcal F}(\Pi^j, \varrho^j)$ is the right-hand side of the integrated algebraic equation (\ref{doplnky-c-eq-ro-integral}). The operator 
${\mathcal T}(\Pi^j, \varrho^j)$ is the transport equation solver given by 
the right-hand side of (\ref{doplnky-c-convective-discrete}) and
${\mathcal A}={\mathcal A}(\Pi^j, \varrho^j)$ is a tridiagonal matrix with coefficients given corresponding to (\ref{findif}). At each time level $\tau_j, j=1, ..., m$, the above system can be solved approximately by means of successive iterations procedure. Given a discrete solution $\Pi^{j-1}$, we start up iterations by defining  $\Pi^{j,0} = \Pi^{j-1}, \varrho^{j,0} = \varrho^{j-1}$. Then the $(p+1)$-th approximation of $\Pi^j$ and $\varrho^j$ is obtained as a solution to the system:
\[
\varrho^{j,p+1} = {\mathcal F}(\Pi^{j,p}, \varrho^{j,p}),\quad
\Pi^{j-\onehalf, p+1} ={\mathcal T}(\Pi^{j,p}, \varrho^{j,p+1}),
\]
\begin{equation}
{\mathcal A}(\Pi^{j,p}, \varrho^{j,p+1}) \Pi^{j,p+1} = \Pi^{j-\onehalf, p+1} \,.
\label{doplnky-c-abstract-iter}
\end{equation}
We repeat the procedure for $p=0,1 ..., p_{max}$, until the prescribed tolerance is achieved. 

\begin{figure}
\sidecaption
\includegraphics[scale=.6]{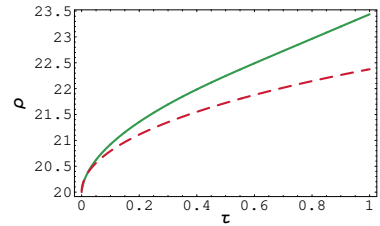}
\caption{
A comparison of $\rho(\tau)=S_f(T-\tau)$ (solid line) for the Barles and Soner model with $a=0.15$ and for the Black-Scholes equation, i.e. $a=0$.}
\label{fig-barles}       
\end{figure}

At the end of this section, we present a numerical example of approximation of the early exercise boundary for the Barles and Soner model by means of a solution to the transformed system of equations. In this model the volatility is given by expression (\ref{doplnky-c-barles}). A discrete solution pair $(\Pi,\varrho)$ has been computed by our iterative algorithm for the  model parameters: $E=10, T=1$ (one year), $r=0.1$ (10\% p.a) , $q=0.05$ (5\% p.a.) and  $\hat\sigma=0.2$. As for the numerical parameters, we chose $n = 750$ spatial points and $m = 225000$ time discretization steps. The step $k=T/m$ represents 140 seconds in the real time scale. In order to achieve the precision $10^{-7}$ we used $p_{max}=6$ micro-iterates in (\ref{doplnky-c-abstract-iter}). A graphical plot of the early exercise boundary $\varrho(\tau)=S_f(T-\tau)$ is shown in Fig.~\ref{fig-barles}. Taking a positive value of the risk aversion coefficient $a=0.15$ resulted in a substantial increase of the free boundary position $\varrho(\tau)$ in comparison to the linear Black--Scholes equation with constant volatility $\sigma=\hat\sigma$. Notice that the Barles and Soner model for $a=0$ coincides with the linear Black--Scholes model with constant volatility. 

%
\bibliographystyle{spmpsci}
\bibliography{p163_ref}  
%



\end{document}